\documentclass[conference]{IEEEtran}
\IEEEoverridecommandlockouts
\IEEEpubid{\makebox[\columnwidth]{979-8-3503-7240-3/24/\$31.00 \copyright 2024 IEEE \hfill}
\hspace{\columnsep}\makebox[\columnwidth]{ }}
\usepackage{cite}
\usepackage{amsmath,amssymb,amsfonts}
\usepackage{graphicx}
\usepackage{textcomp}
\usepackage{xcolor}
\usepackage{algorithm}
\usepackage{algpseudocode}
\usepackage{svg}
\usepackage{tikz}
\usetikzlibrary{shapes.geometric, arrows}
\usepackage{subcaption}
\PassOptionsToPackage{hyphens}{url}\usepackage[hidelinks]{hyperref}

\def\BibTeX{{\rm B\kern-.05em{\sc i\kern-.025em b}\kern-.08em
    T\kern-.1667em\lower.7ex\hbox{E}\kern-.125emX}}
\begin{document}

\title{Cut-set and Stability Constrained Optimal Power Flow for Resilient Operation During Wildfires
\thanks{This work was supported in part by  the National Science Foundation (NSF) grant under Award ECCS-2132904.}
}

\author{\IEEEauthorblockN{Satyaprajna Sahoo, Anamitra Pal}
\IEEEauthorblockA{\textit{School of Electrical, Computer and Energy Engineering} \\
\textit{Arizona State University}\\
Tempe, United States \\
sssahoo2@asu.edu, anamitra.pal@asu.edu}
% \and
% \IEEEauthorblockN{Anamitra Pal}
% \IEEEauthorblockA{\textit{School of Electrical, Computer and Energy Engineering} \\
% \textit{Arizona State University}\\
% Tempe, United States \\
% anamitra.pal@asu.edu}
}
\maketitle
% \vspace{-10pt}
\begin{abstract}
% % Wildfires
% % % , especially power system-related wildfires 
% % pose a significant challenge to modern power grid operation.
% % management. 
% % (raw) 
% % Currently, the risk of grid infrastructure of being affected by wildfires have been mitigated by public safety power shutoffs (PSPS), which lead to extended periods of power loss. Conversely, insufficient action or (defects in working) of infrastructure lead to transient instabilities, in the worst form of which leading to cascading outages, or blackouts.
% Wildfires pose two challenges for power utilities: they must
% % to reliable and resilient power system operation: 
% (a) minimize the risk of starting wildfires, and (b) operate the grid in a secure manner during ongoing wildfires.
% The focus of this paper is on the latter. 
Resilient operation of the power system during ongoing
% under active 
wildfires is challenging because of the uncertain ways in which the fires impact the electric power infrastructure.
For example, wildfires near energized power lines cause arc-faults to occur in rapid succession; such frequent faults can trip multiple grid assets due to dynamic instability. 
% which make it difficult to operate the grid in a reliable and resilient manner.
% Such occurrences can trip multiple assets in the grid.
% grid, making contingency analysis and resilient operation under active wildfire situations a challenging problem.
Conventional contingency analysis tools are not equipped to handle such phenomena.
% and are generally known to trip multiple assets in the grid, making conventional contingency analysis and subsequent operation under wildfire risks challenging.
% Resilient operation of the power system during ongoing wildfires is challenging because of the uncertain ways in which the fires impact the electric power infrastructure (multiple arc-faults, complete melt-down).
% electric grid,
% % power infrastructure, 
% mainly 
% transmission lines (multiple arc-faults to eventual melt-down). 
% \textcolor{red}{Currently, the impact of wildfire induced instabilities to the grid has been limited to improving general resilience, or hardening the grid infrastructure. Concurrently, the focus has been mainly on preventing grid-initiated fires, or improving the grid resilience as a whole. However, analysis of operational improvements that can be made during active wildfire scenarios is relatively unexplored.} 
To address this challenging problem, we propose
% In this paper, we propose 
a novel cut-set and stability-constrained optimal power flow (CSCOPF) that
ensures secure, stable, and economic power system operation through an advanced contingency analysis formulation which
quickly detects and mitigates both static and dynamic insecurities as wildfires progress through a region.
% given the risk of active wildfires in a region. 
% A ``feasibility test" (FT) algorithm that exhaustively desaturates overloaded cut-sets to prevent cascading line outages, and a data-driven transient stability analyzer that predicts the correction factors for eliminating generator trips is integrated with the OPF problem.
The CSCOPF achieves its objective by integrating a ``feasibility test" algorithm that exhaustively desaturates overloaded cut-sets to prevent cascading line outages and a data-driven transient stability analyzer that predicts the correction factors for eliminating generator trips, with the optimal power flow formulation.
% transient instabilities.
% A data-driven transient stability analyzer is first developed that predicts the correction factors for eliminating transient instability.
% Then, it is combined with a Feasibility Test (FT) algorithm that quickly desaturates overloaded cut-sets to prevent cascading outages.
% This regressor is included with the Feasibility Test (FT) algorithm that quickly identifies saturated cut-sets, preventing cascading line outages. 
% The proposed model considers the possibility of
% % includes 
% generation rescheduling as well as load shed. 
% to retain flexibility in multi-horizon dispatches. 
% The solution is tested on the IEEE 118-bus system and a renewable-rich 240-bus system. 
The results obtained using the IEEE 118-bus system
% and a renewable-rich 240-bus system
indicate that the proposed approach alleviates
% and it is found to alleviate 
vulnerability of the system to active wildfires while simultaneously minimizing 
% additional 
operational cost.

\end{abstract}

% \vspace{0.5em}

\section{Introduction}

\IEEEPARstart{I}{n} recent years, the prevalence of wildfires has surged, posing monumental challenges for 
% first responders/emergency services, forest officials, as well as 
electric power utilities. 
% While the impacts on emergency services and forest officials are easy to understand, the interaction between wildfires and the electric power infrastructure is more complex.
On one side, they have been blamed for starting devastating wildfires, which have even culminated in some utilities declaring bankruptcy \cite{balaraman2020wildfires}. 
On the other side, they are expected to maintain secure and stable grid operation in presence of active wildfires to support other critical services.
% The first aspect has been the focus of considerable research in the past five years \cite{10164135,9840510,9305959,9220164}. 
The focus of this paper is on the latter, namely, \textit{ensuring resilient power system operation when a progressing wildfire is projected to intersect a power transmission corridor}.

An analysis of major wildfire-induced power system interruptions has identified \textit{cascading outages resulting from overloaded assets}, \textit{frequent arc-faults}, and/or \textit{preemptive disconnection of power system equipment}
% mis-operations of the protection system 
as the primary causes (for the interruptions) \cite{9737413,wu2016space}.
% \cite{daochun2015review,wu2016space}.
% primary cause(s) (for the interruptions).
% \begin{itemize}
%     \item 
Overloads occur because the heat from the fires affect the power lines in their vicinity resulting in lowering of the conductor's current carrying capacity \cite{9677975}. This lowering may then cause bottlenecks to appear in other parts of the system. 
The pioneering work on this topic was done by \cite{CHOOBINEH201520}.
% The pioneering work on this topic was done by \cite{CHOOBINEH201520}, and the papers that cited it. 
% However, these papers
However, \cite{CHOOBINEH201520} and the papers that cited it, 
% % % (such as \cite{9693217}), 
did not consider the impacts of fire-induced faults on dynamic stability of the power system.
% \item 
During the 2016 Blue Cut Fire, as the fire approached a corridor of three 500kV and two 287kV transmission lines, \textit{15 arc-faults occurred in a short period of time}
% , ultimately resulting in the loss of 1.2GW of solar generation 
\cite{nerc1200}. 
Other instances of system instabilities caused by fire-induced arc-faults can be found in \cite{8620983,daochun2015review,wu2016space}.
% Other instances of system instabilities caused by fire-induced faults can be found in \cite{8620983,daochun2015review,wu2016space,9061890}.
Wildfire induced arc-faults are unique as they occur multiple times within a few seconds
% within a short time-period 
\cite{9449670}. Conventional contingency analysis tools that usually deal with \textit{one fault occurring on a line/lines}, are not equipped to handle such phenomena.
At the same time, preemptively disconnecting lines in advance and over wide regions to protect them from \textit{future} arc-faults bears a very high social cost.
During the 2019 California fires, preemptive actions by the local utility left 
% over 940,000 houses and businesses without electricity which impacted 
more than 2.7 million people without electricity \cite{Columbia}.
% During the 2019 California fires, preemptive actions by the local utility left 
% % over 940,000 houses and businesses without electricity which impacted 
% more than 2.7 million people without electricity \cite{Columbia}.
% \end{itemize}
% the subsequent impact of the sudden disconnection on grid operation was not investigated. 
% The severity of arc-faults can be realized from the fact that during the Blue Cut Fire, \textit{as the fire approached a corridor of three 500kV and two 287kV transmission lines, 15 line faults occurred in a short period of time} \cite{nerc1200}; also see Fig. \ref{fig:wildfirefault}.
% Examples of protection system mis-operations include erroneous frequency tripping (Blue Cut Fire \cite{nerc1200}) and momentary cessation (Canyon 2 Fire \cite{nerc900}).
% In summary,
% Since wildfire risks have increased over wide regions, 
% for extended periods of time, 
Since areas impacted by wildfires have grown considerably in the last decade,
power utilities must operate their systems till the last-minute while also considering
% one must consider 
static security (to protect against asset overloads) and dynamic stability (to minimize impact of frequent arc-faults) \cite{shrestha2021frequency}.

\begin{figure}
 % \vspace{5em}
	\centering
	\includegraphics[width=0.485\textwidth]{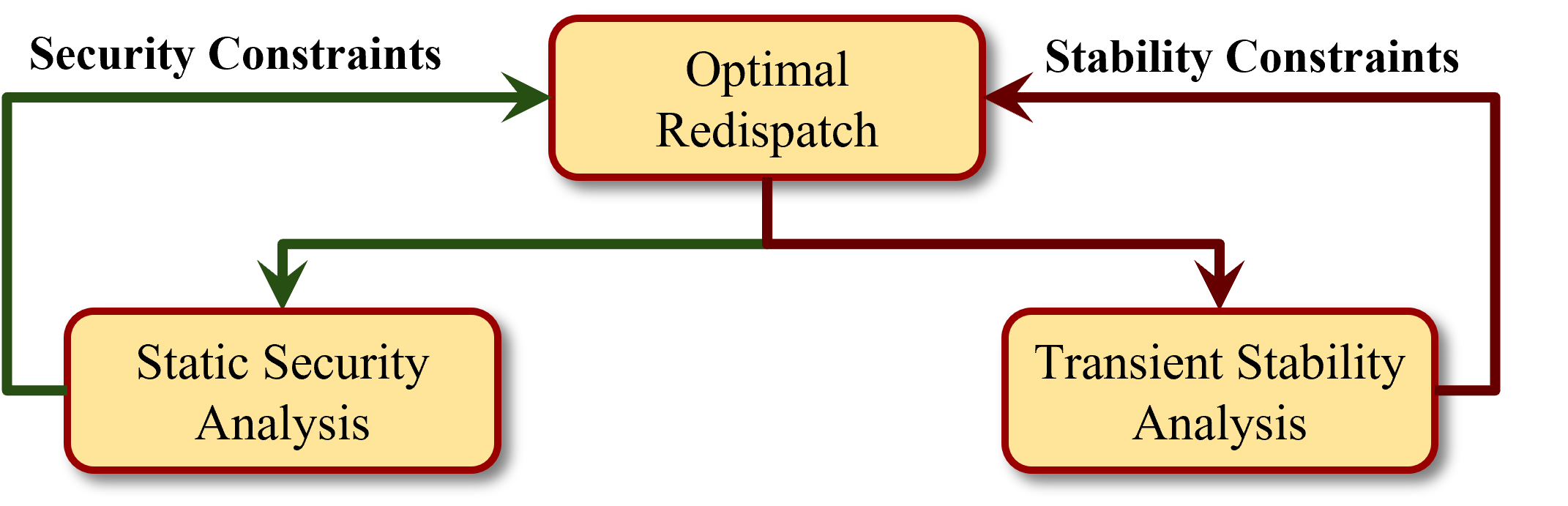}
	\caption{CSCOPF overview}
	\label{fig:overview}
\end{figure}

In this paper, we introduce a novel cut-set and stability-constrained optimal power flow (CSCOPF) 
to conduct a comprehensive contingency analysis for active wildfire scenarios, 
and 
% to 
ensure resilient power system operation upon emergence of such scenarios.
% active wildfire situations.
A cut-set is a set of lines, which if tripped, would create disjoint islands in the network.
Therefore, saturated/overloaded cut-sets are the most vulnerable interconnections of the system 
% because they have limited power transfer capability 
\cite{biswas2020graph}.
For ensuring static security, we leverage the ``feasibility test" (FT) algorithm developed in \cite{biswas2021mitigation} to \textit{exhaustively} protect the system against saturated cut-sets; we also ensure security against $\mathrm{N-1}$ branch overloads.
% as well as branch overloads.
% , which is designed to be entirely convex \textcolor{red}{verify}. 
Dynamic stability is ensured through a data-driven transient stability constraint prediction (TSCP) algorithm that estimates the required transient stability correction factor (TSCF) while accounting for load uncertainties.
The outcomes of the two algorithms are added as constraints to the optimal power flow (OPF) formulation (see Fig. \ref{fig:overview}).
% A brief overview of the CSCOPF is provided in Fig.
% % integration of the contingency analysis with the OPF problem is given in figure
% \ref{fig:overview}. 
The OPF, modeled as an optimal redispatch problem, is run iteratively until all violations are addressed.
% the contingency analysis tools do not flag 
% no violations are found.
% Central to our proposal is a data-driven Transient Stability Constraint Prediction (TSCP) algorithm to effectively alleviate all generator outages. 
% The regressor predicts the Transient Stability Correction Factor (TSCF) based on forecasted loading conditions, which represent significant uncertainties within the system. 
% Further, this solution is integrated with a Feasibility Test (FT) algorithm developed in \cite{biswas2020graph} which exhaustively identifies and mitigates all potential cascading line outages in the system. 
% For the effective deployment of the CSCOPF formulation, a two-stage optimization approach is followed. The day-ahead stage encompasses a single-stage optimization, while the real-time phase requires a more restricted multi-stage optimization. To ensure flexibility in implementation, the security and stability constraints are enforced using generation rescheduling, renewable resource curtailment, as well as load shedding.
% To ensure flexibility in implementation, the security and stability constraints are enforced using generation rescheduling as well as load shedding.
A case-study conducted using the IEEE 118-bus system 
% and a renewable-rich 240-bus Western Electricity Coordinating Council (WECC) system
demonstrates that the proposed approach is able to alleviate cascading outages due to static and dynamic insecurities with minimal increase in operational cost.

\section{Problem Scope and Solution Approach}
% Analysis of power lines affected by wildfires show that the fires can cause \textit{multiple arc faults within a short time-period},
% % period.
% % % faults begin as multiple arcing events, that eventually lead to a flash-over. 
% % Each arc fault negatively impacts the stability of the system, and such faults occurring frequently 
% which may lead to generator desynchronization, and \textit{permanent line outages}, which may cause bottlenecks to appear in other parts of the system \cite{ma2020real}. 

Wildfires and their interaction with the electric power grid
% built-infrastructure 
constitutes a multi-faceted problem because of the following reasons: (a) the
% \textcolor{red}{The} 
spatio-temporal process of wildfire spread is based on local climate and geography (topography, vegetation, wind); (b) breakdown mechanisms of the air gap varies with time, location, and wildfire proximity and intensity; (c) 
outcomes can range from multiple arcing events to a complete line melt-down (permanent outage)
% faults begin as multiple arcing events and eventually culminate in a line meltdown (permanent fault)
% faults begin as multiple arcing events, followed by a flash-over, and eventually line meltdown (permanent outage) 
\cite{ma2020real,daochun2015review,8620983}.
We now specify the problem scope in the context of these facets.

A variety of tools already exist for tracking wildfire spread over different geographical regions (e.g., FlamMap \cite{flammap}); hence,
% Therefore, 
tracking of wildfires is outside the scope of this paper.
% Therefore, this study assumes that knowledge of when an ongoing wildfire will get into the security buffer of power lines is known a priori.
% (e.g., a day in advance).
Without knowledge of the local environmental conditions around a transmission line when a fire is nearby, it is not possible to determine the air quality and/or the type of event that might occur (frequent multiple arc-faults or permanent outage). 
In this regard, one strategy could be to assume that all power lines located in an active wildfire area are preemptively de-energized, and then solve an optimal generation redispatch problem in presence of topology changes \cite{9693217}. However, as explained in \cite{9737413}, such a strategy may not be optimal from a socio-techno-economic perspective.

In this paper, we consider both types of outcomes, namely, multiple arc-faults as well as permanent line outages. 
% To adequately prepare the system for such extreme events, two contingency analysis methods are proposed.
To prepare the system for such eventualities, we develop two contingency analysis methods.
% Two contingency analysis methods are described below to prepare the system for such eventualities. 
Although these methods can be used for any extreme event/multi-asset contingency, 
% this paper focuses specifically on the unique attributes of wildfires.
they are implemented (see Section \ref{Implementation})
% in this paper 
considering the unique attributes of wildfires.
% To note, the following analysis methods and formulation can be used for any extreme event/multi-asset contingency, however this paper focuses specifically on wildfires.
Lastly, some articles have focused on enhancing the resilience of the power grid against wildfires by system hardening, better asset management, and/or optimally allocating fire-extinguishing resources \cite{8244267,9677975,9409078,10032578}.
% 9220164,9409078,10032578}. 
These could be deemed complementary to the scope of this paper.

\subsection{Static Security using Cut-set Analysis}
A saturated cut-set
% By monitoring power flow across cut-sets, security criteria across multiple loss of lines can be strengthened by detecting saturated cut-sets, which 
is 
% defined as 
a cut-set whose aggregate power flow exceeds the limits of the constituent lines, as shown below: 
% for a cut-set having $k$ lines:
% \begin{equation}
% % \label{eq:saturated_cut-set}
% %     \sum_{e=1}^k f_e > \sum_{e=1}^k f^{\mathrm{max}}_e
% % \end{equation}
\begin{equation}
\label{eq:saturated_cut-set}
    \sum_{\forall e \in K_{\mathrm{crit}}} f_e > \sum_{\forall e \in K_{\mathrm{crit}}} f^{\mathrm{max}}_e
\end{equation}
% \nomenclature{\(f_e\)}{Active power flow in line $e$}
where,
% In \eqref{eq:saturated_cut-set}, 
$f_e$ denotes the flow in the $e^{th}$ line, $f^{\mathrm{max}}_e$ is the maximum flow allowed through that line, and $K_{\mathrm{crit}}$ is a saturated cut-set.
A transmission corridor of the power system is likely to be a part of one or more cut-sets.
% A fast and scalable algorithm called Feasibility Test (FT) is used to exhaustively identify and alleviate all saturated cut-sets for a given contingency \cite{biswas2020graph}.
The FT algorithm, which takes power injections as its inputs \cite{biswas2020graph,sahoo2023data}, quickly identifies and alleviates all saturated cut-sets for a given contingency, and is mathematically expressed as,
% as shown below,
\begin{equation}
\label{eq:model_cut_set_constraint}
    \sum_{\forall e \in K_{\mathrm{crit}}} \Delta f_e \leq - \Delta P_{K_{\mathrm{crit}}} \quad  \forall K_{\mathrm{crit}} \in \kappa_{\mathrm{crit}}
\end{equation}
% \nomenclature{\(\Delta P_{K_{crit}} \)}{Transfer margin of cut-set $K_{crit}$}
 % \nomenclature{\(K_{crit} \)}{Identified critical cut-set}
% \nomenclature{\(\kappa_{crit} \)}{Set of all identified critical cut-sets}
where, 
$\kappa_{\mathrm{crit}}$ is the set of all identified cut-sets, and 
$\Delta P_{K_{\mathrm{crit}}}$ is the identified transfer margin, which is the overall amount of power that must be reduced across the lines in the cut-set to 
% achieve security.
eliminate overloading.
% \autoref{eq:model_cut_set_constraint} shows the formulation of a constraint that is generated when a particular cut-set is saturated for a contingency. 
% While the Feasibility Test is able to assess the effect of extreme events on transmission lines, it does not consider the transient stability of multiple successive contingencies on the generators. 

% \vspace{-5pt}
\subsection{Transient Stability using Machine Learning} 
Transient stability assessment and control evaluates the capability of the power system to maintain synchronism after a large disturbance, such as multiple frequent arc-faults. 
% It is vital because loss of synchronism in generators leads to tripping, causing blackouts. 
% Various variables
Different attributes of the system, including system configuration, loading conditions, and fault location, influence the transient stability assessment. 
% The most popular TSA method is 
In this paper, we focus on the
rotor angle stability, which is a key factor in determining transient stability in presence of faults \cite{9290096}.
The rotor angle stability is
quantified by the transient stability index ($\mathrm{TSI}$) shown below, where $\delta_{\mathrm{max}}$ is the maximum difference in the sorted rotor angles of two consecutive machines.
% Transient stability is assessed from the difference in the maximum rotor angle ($\delta_{\mathrm{max}}$) between two machines. It results in the transient stability index ($\mathrm{TSI}$) shown below:
% Cascading outages caused by wildfires possess the ability to cause rotor angle instabilities in the generators. Stability is assessed using the maximum difference of the rotor angles between two consecutive machines $(\delta_{max})$, which yields an index
\begin{equation}
    \label{eq:eeac1}
    \mathrm{TSI} = \frac{360 - \delta_{\mathrm{max}}}{360 + \delta_{\mathrm{max}}} \times 100
\end{equation}

% \nomenclature{\(\delta_i \)}{Rotor angle of generator $i$}

The system is stable if $\mathrm{TSI}>0$, and unstable otherwise.
% The transient stability index (TSI) reflects system stability, and the system is stable when $\mathrm{TSI}>0$, and unstable when $\mathrm{TSI}<0$. 
% For an unstable contingency, the total generation can be classified into stable and unstable generators, with the stable generators being those whose rotor angles are below $\delta_{\mathrm{max}}$.
For a contingency that causes $\mathrm{TSI}$ to become negative, the total generation is split into two groups, one of which is composed of the critical machines (CM) and the other is composed of the non-critical machines (NM).
A machine (i.e., a generator) is deemed critical when it swings away from the rest of the machines. Note that there can be more than one machine that is identified as critical for a given contingency. 

% To calculate a stability index for IBRs, we use the phase angle of the current that the inverter injects into the grid relative to the grid voltage. This angle is controlled electronically and determines the active and reactive power that the inverter supplies to the grid.
% \begin{equation}
%     TSI_{IBR} = \frac{360 - \phi_{max}}{360 + \phi_{max}} \times 100
% \end{equation}
%  where $\phi_{max}$ is the maximum difference between the current angles of two consecutive generators.\\

Given a contingency, transient stability control can be
% is usually 
done by the
single machine equivalent (SIME) method \cite{zhang1998sime}, which 
% is a hybrid TSA method that 
utilizes the equal area criterion (EAC) and multiple time domain simulations (TDSs) per contingency. 
% When an instability occurs, it can be corrected using the integrated extended equal area criterion (IEEAC).
When an instability occurs in a multi-machine system, the integrated extended equal area criterion (IEEAC) can be employed to correct for the instability.
% \cite{zhang1997sime}.
% Given a contingency, instability correction is done using the Extended Equal Area Criterion (EEAC) \cite{zhang1997sime}. 
The IEEAC 
% of the single machine equivalent (SIME) of a multi-machine system 
stipulates transferring power (denoted by $P_{tr}$) from the unstable generators to the stable generators, as shown  below \cite{7395386}:
\begin{equation} 
    \label{equation:Pshift}
    \Delta P_{tr} \geq \left(\frac{-\eta_{us} + \epsilon}{\tau} \right).\left(\frac{M}{M_{\mathrm{CM}}} + \frac{M}{M_{\mathrm{NM}}}\right)^{-1}
\end{equation} 
% \nomenclature{\(M_i \)}{Inertia constant of generator i}
where, $M, M_{\mathrm{CM}}, M_{\mathrm{NM}}$ are the one machine infinite bus (OMIB) inertia coefficients of the whole system, critical machines ($\mathrm{CM}$), and non-critical machines ($\mathrm{NM}$),
% OMIB inertia coefficient of the CMs, and OMIB inertia coefficient of the NMs,
respectively, 
% $\Delta  P_{tr}$ is 
% referred to as 
% the transient stability correction factor (TSCF), 
$\eta_{us}$ and $\epsilon$ are the unstable value and desired value of the transient stability margin, and $\tau$ is a sensitivity factor.
% . 
% The relationship between the transient stability margin and the one machine mechanical power has been found to be quasi-linear \cite{zhang1997sime} with $\tau$ being the sensitivity factor.
$\Delta P_{tr}$, which is the TSCF, can be used to create the transient stability constraint as shown below:
\begin{equation}
\label{eq:trans_stab_const}
    \sum_{\forall i \in \mathrm{CM}} \Delta p_{i} \leq - \Delta P_{tr}
\end{equation}

Now, the calculation of $\Delta P_{tr}$ depends on multiple TDSs for a single contingency, which is computationally expensive to do in real-time.
% Further, the load is not known precisely, which makes any calculation of $\Delta P_{tr}$ in advance inaccurate to a certain degree. 
Furthermore, one must also consider the variability of the loads.
To account for both of these factors, we exploit the quasi-linear relationship between the pre-contingency one machine mechanical power and the transient stability margin \cite{7395386}.
% In \eqref{equation:Pshift}, the relationship between the pre-contingency one machine mechanical power and the transient stability margin is quasi-linear. This can be extended to posit a linear relationship between the transfer margin and the pre-contingency loading condition as well, since the pre-contingency mechanical power is linearly related to the loads.
Specifically, we posit that \textit{a linear relationship exists between the TSCF
% transfer margin
and the pre-contingency loading condition}. This is because 
% since 
the pre-contingency mechanical power is linearly related to the loads ($l$) as $ \sum_{i \in G} p^{m}_{i} = \sum_{i \in G} p^{e}_{i} = \sum_{j \in L} l_j $, where $p^{m}_{i}$ and $p^{e}_{i}$ denote mechanical and electrical power, respectively, of the $i^{th}$ generator.
Consequently, we formulate a linear regression (LR) model with a mean squared error (MSE) loss function
% As such,
% To account for this uncertainty, 
% a linear regression (LR) model is formulated below 
that estimates the required TSCF for a given
% forecasted 
loading condition, as shown below:
% \textcolor{gray}{
%     \begin{itemize}
%         \item A brief description of the theory of SIME, with particular focus on the quasi-linear relationship between the stability index and real generated power.
%         \item Stability index leads to P-shift, the constraint for corrective action
%         \item However, DER's have no rotor angle, hence no stability index. Solution?
%         \item Since DER's are emulated to resemble SG behaviour, the relationship between Pshift and Pgen still holds.
%         \item Further, Pgen is linearly correlated with Pload, and this relationship can be learned.
%     \end{itemize}
% }
% \subsection{Data-driven model for transient stability correction in the face of load uncertainty}
% For transient stability corrective action in IBR rich systems, we leverage the relationship between $\Delta P_C$ and $P_e$ as
% From EEAC, we can express $\Delta P_C$ in terms of $P_{ei}^0$ as 
% \begin{equation}
%     \Delta P_C = f(P_e)
% \end{equation}
% where, $f()$ can be derived by combining equations \eqref{equation:Pshift} through \eqref{equation:OMIB_inertia}. To account for load uncertainty, we extend the correlation by substituting $P_e$ with the load $P_L$ as
% % \begin{equation}
% %     \sum_{i \in L} P_l = \sum_{j \in G} P_{ei}
% % \end{equation}
% \begin{equation}
%     \Delta P_C = f(g(P_L))
% \end{equation}
% where, the function $g()$ is derived through the energy balance equation $\sum_{i \in L} P_l = \sum_{j \in \{G,R\}} P_{i}$. The function $f \circ g$ is modelled via a linear regression (LR) model, with a mean squared error (MSE) loss
\begin{equation}
\label{eq:lin_reg_forward}
    \Delta \hat{P}_{tr} = \sum_{j \in L} \theta_j l_j + \theta_0  
    = \Upsilon(l_j)
 \end{equation}
 % \vspace{-6pt}
 \begin{equation}
     J(\theta) = \frac{1}{k} \sum_{i=1}^k (\Delta \hat{P}_{tr_i} - \Delta {P}_{tr_i})^2
 \end{equation}
% \nomenclature{\(M_i \)}{Inertia constant of generator i}
% \nomenclature{\(M_i \)}{Inertia constant of generator i}
where, $\theta$ are the weights, $J(\theta)$ is the loss function, and $k$ is the batch size. This LR-based model for estimating the TSCF is referred to as the TSCP algorithm, and is denoted by $\Upsilon$.
% where, $\Delta \hat{P}_C$ is the estimated value of the transfer margin, and $\theta$ are the trainable weights of the linear regression model, whose loss for a training batch of size $m$ is $J(\theta)$. 
% One of the reasons for choosing a linear regression model was its convex formulation, which can be derived by calculating the Hessian ($H$) of the loss with respect to the weights $\theta$

% \begin{equation}
%     H = \triangledown ^2 J(\theta) = \frac{1}{k} X^T X
% \end{equation}
% where $X$ is the input matrix ($P_L$) for a single batch. The Hessian is positive semi-definite, implying convexity. 

% In the regression model (denoted as $\Upsilon$), $\theta$ are the weights, and $J(\theta)$ is the loss function, which is a standard mean squared loss (MSE). 

\vspace{0.5em}

\section{CSCOPF Formulation and Implementation}
\label{Section2}

\subsection{Objective and Constraints}
\label{sec:2b}

% CSCOPF is modeled as an optimal redispatch problem to alleviate overloaded cut-set and transient stability violations.
% % identified for a given contingency. 
% The generator costs are modeled as 
% % quadratic functions as 
% shown below,
% % The objective a part of the optimal power flow problem, where each generator is expected to have a quadratic cost $(F_i)$  w.r.t it's electrical output ($P_e$). 
% We start by modeling the generator costs as shown below,

% \begin{equation}
%     \label{eq:gen_cost_base}
%     F_i (p_{i}) =  {a_i + b_i  p_{i} +  c_i (p_{i})^2 }
% \end{equation}
% where, $a,b,c$ are the cost coefficients of the $i^{th}$ generator. To derive the cost of generation change, \eqref{eq:gen_cost_base} is expressed as

% CSCOPF is modeled using a warm start model, since the solution is to be implemented over the economic dispatch if the wildfire manifests. 
CSCOPF is modeled as an optimal redispatch problem to alleviate saturated cut-set, branch overloads, and transient stability violations.
The cost ($F$) of generation dispatch considering a quadratic cost curve, can be expressed as:
\begin{equation}
    \label{eq:gen_cost_base}
    F_i (p_{i}) =  {a_i + b_i  p_{i} +  c_i p_{i}^2 }
\end{equation}
where, $a_i,b_i,c_i$ are the fixed,
% no-load, 
linear, and quadratic cost coefficients of the $i^{th}$ generator; we also dropped $e$ from the superscript of $p^{e}_{i}$ to avoid notational clutter.
To derive the cost of generation change, \eqref{eq:gen_cost_base} is expressed as
% \begin{equation}
% \label{eq6}
% \begin{aligned}
% \Delta F_i (\Delta p_{i}) =  c_i (\Delta p_{i})^2 + ( b_i + 2 c_i p^0_{i})\Delta p_{i}\\
% \end{aligned}
% \end{equation}
\begin{equation}
\label{eq6}
\begin{aligned}
\Delta F_i (\Delta p_{i}) &=   \{a_i + b_i  p_{i} +  c_i p_{i}^2 \}  - \{a_i + b_i  p^0_{i} +  c_i (p^0_{i})^2 \} \\
 &  = c_i \Delta p_{i}^2 + ( b_i + 2 c_i p^0_{i})\Delta p_{i} 
\end{aligned}
\hspace{-4pt}
\end{equation}
where, the superscript $^{0}$ in \eqref{eq6} refers to the pre-contingency status.
% However, shifting generation may not be a feasible solution for \textit{every} contingency, as it may cause branch or cut-set overloads in another area of the system.
% % Morever, synchronous generators generally have ramp-rate limits, which limit flexibility.
% In such cases, load is shed to alleviate grid vulnerability. Thus, the overall objective is written as
Including load shed ($\Delta l$), the overall objective becomes minimizing redispatch as well as
% is to minimize redispatch cost while minimizing 
load shed, as shown below:
\begin{equation}
    \label{eq:new_obj}
    \begin{aligned}
    \mathrm{min} \quad &  \sum_{\forall i \in G} (c_i \Delta p_{i}^2 + d_i \Delta p_{i}) + \sum_{\forall j \in L} (m_j \Delta l_j)
    \end{aligned}
\end{equation}
% \begin{equation}
%     \begin{aligned}
%     \mathrm{min} \quad &  \sum_{\forall i \in G} (c_i \Delta P_{ei}^2 + d_i \Delta P_{ei}) + \sum_{\forall j \in L} (m_j \Delta L_j)\\
%      \quad & + \sum_{ \forall l \in R} (n_l \Delta P_{rl}) + \sum_{G_d} u_i a_i\\
%     \end{aligned}
% \end{equation}
% \begin{equation}
%     \begin{aligned}
% G^s: \quad & \text{set of all generators}\\
% L^s: \quad & \text{set of all loads} \\
% R^s: \quad & \text{set of all renewable generators}\\
% G_d^s: \quad & \text{set of all inactive generators} \\
% \Delta G_e: \quad & \text{active power output of SG i}\\
% \Delta L: \quad & \text{Load shed}\\
% \Delta G_r: \quad & \text{renewable curtailment of gen i}\\
% a_i, c_i, d_i: \quad & \text{cost coefficients of SG i}\\
% m_j: \quad & \text{Cost of load shed}\\
% n_l: \quad & \text{cost of renewable curtailment}\\
% u_i: \quad & \text{binary commitment status of SG i }\\
%     \end{aligned}
% \end{equation}
% \begin{equation}
%     \begin{aligned}
% \rho: \quad & \text{Ramp rate limit}\\
% PTDF: \quad & \text{Power transfer distribution factor}\\
% f: \quad & \text{Branch active power flow}\\
%     \end{aligned}
% \end{equation}
where, $d_i = (b_i + 2 c_i p^0_{i})$, $m_j$ is the cost of shedding load $j$, and $G$ and $L$ are the sets of generators and loads in the network.
% The objective is subject to the following constraints:
The coefficient $m_j$ is typically chosen to be higher than the generation costs, to disincentivize 
% power outage. 
load shedding.
% However, depending upon the situation, they are changed to produce more practical results.

\subsection{Variable Limit Constraints}
The generation rescheduling 
% ($\Delta P_{e}$) 
and load shed 
% ($\Delta L$) 
are limited by the following equations, where the superscripts $\mathrm{max}$ and $\mathrm{min}$ refer to the maximum and minimum values of the scripted variables, respectively.
% \begin{equation}
%     u_i . (P_{i}^{min} - P_{i}^0) \geq \Delta P_{i} \geq u_i . (P_{i}^{max} - P_{i}^0) \;  \forall i \in \{G,R\}
% \end{equation}
\begin{equation}
    \label{eq:lim_1}
    p_{i}^{\mathrm{min}} - p_{i}^0 \leq \Delta p_{i} \leq p_{i}^{\mathrm{max}} - p_{i}^0 \quad  \forall i \in G
\end{equation}
% \begin{equation}
%     \label{eq:lim_1}
%     p_{e_i}^{\mathrm{min}} - p_{e_i}^b \geq \Delta p_{e_i} \geq p_{e_i}^{\mathrm{max}} - p_{e_i}^b \quad  \forall i \in G
% \end{equation}
\begin{equation}
    \label{eq:lim_2}
    l_j^{\mathrm{min}} - l_j^0 \leq \Delta l_j \leq l_j^{\mathrm{max}} - l_j^0 \quad \forall j \in L
\end{equation}
% In \eqref{eq:lim_1} and \eqref{eq:lim_2}, 

\subsection{Branch Flow Constraints}
The limits on power flows corresponding to the changes in the generation and loads are given by,
\begin{equation}
    \label{eq:lim_3}
\begin{aligned}
      & f_u^{\mathrm{min}} - f_u^0 \leq \sum_{\forall i \in G} \mathrm{PTDF}^r_{u,i} \Delta p_{i} - \sum_{\forall j \in L} \mathrm{PTDF}^r_{u,j} \Delta l_j \\
       & \leq f_u^{\mathrm{max}} - f_u^0 \; \; \; \; \forall u \in B
\end{aligned}
\end{equation}
% \begin{equation}
%     \label{eq:lim_3}
% \begin{aligned}
%      \quad & f_u^{max} - f_u^b \geq \sum_{\forall i \in G} \mathrm{PTDF}^r_{u,i} \Delta p_{i} - \sum_{\forall j \in L} \mathrm{PTDF}^r_{u,j} \Delta l_j \\
%       \quad & \geq f_u^{min} - f_u^b   \quad \forall u \in B\\
% \end{aligned}
% \end{equation}
% \begin{equation}
%     \label{eq:lim_4}
%     \sum_{\forall i \in \{G\}} \mathrm{PTDF}^r_{l,i} \Delta P_{e_i} - \sum_{\forall j \in L} \mathrm{PTDF}^r_{l,j} \Delta L_j \geq f_l^{min} - f_l^0 
% \end{equation}
where, $\mathrm{PTDF}^r_{u,i}$ is the power transfer distribution factor of branch $u$ for one unit of power added at bus $i$ and one unit of power withdrawn from reference bus ($r$), $B$ is the set of all branches, and $f_u$ is the active power flowing in $u$. 

% \subsection{Conservation of energy}
\subsection{Power Balance Constraint}
% The aggregate change in generation should equal the total load shed to ensure power balance. This is mathematically expressed as
This is ensured by making the aggregate change in generation equal the total load shed, as shown  below:
\begin{equation}
    \label{eq:lim_5}
    \sum_{\forall i \in G} \Delta p_{i} = \sum_{\forall j \in L} \Delta l_j
\end{equation}

\subsection{Branch Overload Constraints}
The $\mathrm{N-1}$ security criteria must be preserved by the proposed corrective action. This is ensured
% can be maintained 
by using the line outage distribution factor ($\mathrm{LODF}$), as shown below,
\begin{equation}
\label{eq:lim_6}
\begin{aligned}
    \quad & \sum_{\forall i \in G} (\mathrm{PTDF}^r_{u,i} + \mathrm{LODF}_{u,a} \mathrm{PTDF}^r_{a,i}) \Delta p_{i}\\
     - \quad & \sum_{\forall j \in L} (\mathrm{PTDF}^r_{u,j} + \mathrm{LODF}_{u,a} \mathrm{PTDF}^r_{a,j}) \Delta l_j\\
    \leq \quad & f_u^{\mathrm{max}} - f_u^0 + (\mathrm{LODF}_{u,a} f_a^0) \quad \forall u,a \in B, \xi
\end{aligned}
\end{equation}
\begin{equation}
\label{eq:lim_7}
\begin{aligned}
    \quad & \sum_{\forall i \in G} (\mathrm{PTDF}^r_{u,i} + \mathrm{LODF}_{u,a} \mathrm{PTDF}^r_{a,i}) \Delta p_{i} \\
     - \quad & \sum_{\forall j \in L} (\mathrm{PTDF}^r_{u,j} + \mathrm{LODF}_{u,a} \mathrm{PTDF}^r_{a,j}) \Delta l_j\\
    \geq \quad & f_u^{\mathrm{min}} - f_u^0 + (\mathrm{LODF}_{u,a} f_a^0) \quad \forall u,a \in B, \xi
\end{aligned}
\end{equation}
where, $\mathrm{LODF}_{u,k}$ represents the percentage of power in branch $k$ that flows through branch $u$
% $l$ 
if there is an outage of branch $k$, and $\xi$ denotes the contingency set.

\subsection{Cut-set and Transient Stability Constraints}
% % This section of constraints relies on a saturated cut-set alleviation algorithm developed in \cite{biswas2021mitigation}. 
% These constraints are obtained from the FT algorithm 
% % developed in 
% \cite{biswas2021mitigation}.
% % A cut-set is defined as a set of branches that separate the network into distinct islands. 
% It
% % , called Feasibility Test (FT), 
% employs a fast recursive method to exhaustively identify and mitigate all cut-set violations for consecutive $\mathrm{N-1}$ contingencies, and is mathematically written as,
The 
% security constraints include the 
post contingency cut-set constraint is obtained from the FT, as shown below:
\begin{equation}
\label{eq:lim_8}
    \begin{aligned}
        \quad & \sum_{\forall i \in G} (\sum_{\forall u \in K_{\mathrm{crit}}} \mathrm{PTDF}_{u,i}^r) \Delta p_{i} \\
        - \quad & \sum_{\forall j \in L} (\sum_{\forall u \in K_{\mathrm{crit}}} \mathrm{PTDF}_{u,j}^r) \Delta l_j\\
        \leq \quad & - \Delta P_{K_{\mathrm{crit}}} \quad \forall K_{\mathrm{crit}} \in \kappa_{\mathrm{crit}}
    \end{aligned}
\end{equation}

Similarly the transient stability constraint derived from $\Upsilon$ for a defined contingency is given 
% by \eqref{eq:trans_stab_const}.
by,
\begin{equation}
\label{eq:lim_9}
    \sum_{\forall i \in \mathrm{CM}} \Delta p_{i} \leq - \Upsilon(l)
\end{equation}

% Equations \eqref{eq:lim_1}-\eqref{eq:lim_3} impose limits on the optimization variables.
% % The superscript $^0$ refers to the pre-contingency status of the respective variables.
% The power balance equation
% % law of conservation of energy 
% is expressed via \eqref{eq:lim_5}. The post-contingency branch
% % N-1 security 
% overloads are accounted for
% % expressed 
% in \eqref{eq:lim_6}-\eqref{eq:lim_7}. $\mathrm{PTDF}$ and $\mathrm{LODF}$ are the power transfer distribution factors (w.r.t. the reference bus $r$) and line outage distribution factors, respectively. 
% $\xi$ is the contingency list, and $B$ denotes the set of branches.
% % each unstable contingency has its own corresponding constraint.
% Additional post contingency cut-set constraints are modeled in \eqref{eq:lim_8}, where the limit $\Delta P_{K_{\mathrm{crit}}}$ is calculated from the FT algorithm. Finally, \eqref{eq:lim_9} models the transient stability constraint.
Note that the CSCOPF formulation described above is implemented on top of a regular economic dispatch;
% when the wildfire manifests; 
i.e., it uses the prior economic dispatch solution as a \textit{warm start}.
% where, $K_{crit}$ is an identified saturated cut-set with a transfer violation of $\Delta P_{cut}$ 
% % (in MW) 
% in the set $\kappa_{crit}$ of all identified saturated cut-sets. 
% % With this all static security constraints are covered. Now the focus is on the transient stability constraints.
% This completes the formulation of the static security constraints. The transient stability constraints and overall implementation are explained in the next section.
% \section{Transient Stability assessment using (insert variable)}

% Transient stability assessment in conventional multi-machine systems can be broadly classified in three categories: white-box, grey box, and black box models. White box models are models that use measurable system parameters like voltage, frequency, rotor angles etc. and use pre-determined algorithms to calculate a stability index. Grey box models utilize these parameters to create semi determined models (for eg. Lyapunov models) which then give an estimation of stability. Black box models use data driven approaches such as deep neural networks and evolutionary algorithms to estimate stability. 

% \textcolor{gray}{
% \begin{itemize}
%     \item Instead of using the rotor angles, 
%      \item Use Lyapunov based methods
%      \item Use Machine learning
% \end{itemize}
% }

\subsection{Implementation}
\label{Implementation}
The steps that must be followed for implementing the proposed CSCOPF are summarized in Algorithm \ref{alg:cap2}.
% The overall algorithm that must be used to implement the proposed CSCOPF scheme is illustrated in Algorithm \ref{alg:cap2}. 
The implementation occurs over two stages. In the day-ahead stage, contingency analyses is performed for potential wildfire scenarios\footnote{The wildfire spread scenarios can be determined using FlamMap \cite{flammap}.} using SIME to identify instability-inducing contingencies.
% (along with real time contingency analysis). 
Identification of the appropriate contingency set is followed by creating a dataset of potential loading conditions
% which can be 
obtained from historical load data. 
% This data is then used to train the TSCP model, $\Upsilon$, 
% after using SIME 
The TSCP model, $\Upsilon$, is then trained
to calculate the TSCF for the loading conditions and contingencies in the set. 
In real-time, the system is run normally
% can be run economically 
% until the wildfire enters the security buffer of the power lines \cite{8620983}.
until a progressing wildfire is projected to intersect a power transmission corridor.
% except for the few hours when the wildfire manifests,
% until the risk of wildfire is so high, the only other alternative is PSPS, 
Once this happens,
% in which case 
the contingency analysis model is used to generate the cutset and transient stability constraints. Since the CSCOPF constraints are convex\footnote{Equations \eqref{eq:lim_1}-\eqref{eq:lim_9} 
% and  \eqref{eq:trans_stab_const} 
are convex as they are a linear function of the decision variables.}, the optimization can quickly generate an optimal redispatch solution that ensures a secure, stable, and economic
% an appropriate 
response to the unfolding wildfire scenario.
% The algorithm for implementation of the proposed CSCOPF scheme is illustrated in \ref{alg:cap2}. 

% The inputs to CSCOPF would hence include the state estimator and the TDS results. 

\begin{algorithm}
\caption{CSCOPF Implementation}\label{alg:cap2}
\textbf{Input:} Historical load distributions, contingency information ($\xi$), and real-time power injections  \\
\textbf{Output:} Generation redispatch ($\Delta p$), and load shed ($\Delta l$)

% \textbf{Day-ahead:}
% \begin{algorithmic}[1]
% \State Perform random sampling from the load distributions to generate load forecasts 
% \State Use SIME to generate constraint using \eqref{equation:Pshift}
% \State Train $\Upsilon$ using  \eqref{eq:trans_stab_const} and TDSs
% \end{algorithmic}
\textbf{Day-ahead Stage:}
\begin{algorithmic}[1]
\State Perform random sampling from the 
% load distributions to generate load forecast dataset 
historical load distributions to determine potential loading conditions $\Xi$
\State Define empty list of constraints $\Phi$
\For{loading condition
% profile 
in $\Xi$}
    % \State Perform FT using $\xi$
    \State Perform TDS using $\xi$
    \If{violations detected}
        % \State Use FT to generate cut-set constraint using \eqref{eq:model_cut_set_constraint}
        \State Generate transient stability constraint using \eqref{equation:Pshift}
        \State Update $\Phi$
    \EndIf
\EndFor
\State Train $\Upsilon$ using $\Phi$
\end{algorithmic}

\textbf{Real-time Stage:}
% \begin{algorithmic}[1]
% \State Run FT algorithm to generate static security constraints using \eqref{eq:model_cut_set_constraint}
% \State Use trained $\Upsilon$ to generate dynamic stability constraints using \eqref{eq:lin_reg_forward}
% \State Solve \eqref{eq:new_obj} s.t. constraints \eqref{eq:lim_1}-\eqref{eq:lim_9} are satisfied
% \end{algorithmic}
\begin{algorithmic}[1]
\State Define empty list of constraints $\Phi$
\State Obtain real-time 
% system information from state estimator
power injection information using \cite{sahoo2023data}
% real-time state estimation injection data
    \State Generate cut-set constraint using \eqref{eq:lim_8}
    \State Generate transient stability constraint using $\Upsilon$; see \eqref{eq:lim_9}
    \State Update $\Phi$ with constraints \eqref{eq:lim_8} and \eqref{eq:lim_9}
\While{violations detected}

% \While{wildfire is projected to impact a corridor}
    \State CSCOPF: Define objective \eqref{eq:new_obj}, apply constraints
    \Statex $\quad \;$ \eqref{eq:lim_1}-\eqref{eq:lim_7}, and add constraints present in $\Phi$ \label{Step}
    % \State Subject to constraints \eqref{eq:lim_1}-\eqref{eq:lim_7}
    % \State Add constraints present in $\Phi$ to objective \eqref{eq:new_obj}
    \State Solve CSCOPF to get $\Delta p$ and $\Delta l$
    % \State Add additional constraints in $\Phi$
    \State Update dispatch and run power flow
    \State Run FT and single TDS on updated dispatch
    \If {violations detected}
        \State Update $\Phi$ and Go to Step \ref{Step} of Real-time Stage
    \EndIf
\EndWhile
% \While{additional violations detected in step 9}
\end{algorithmic}

\end{algorithm}

% \begin{itemize}
%     \item Preventive formulation, corrective formulation
%     \item Recursive FT with transient stability constraints included
    
% \end{itemize}

% \textcolor{gray}{
% \begin{itemize}
%     \item Potential to add generation uncertainty as well?
% \end{itemize}
% }
% \subsection{Performance guarantee}
% \begin{itemize}
%     \item Derive the mathematical correlation between Pshift and Pload
%     \begin{itemize}
%         \item Correlation between Pshift and stability index: SIME 
%         \item Corr bw stability index and Pgen: SIME
%         \item Corr bw Pgen and Pload: derive
%         \item Point of concern: SIME/OMIB assumptions and their impact
%     \end{itemize}
% \end{itemize}

% \begin{itemize}
%     \item Cite Reetam's papers, formulate the iCA corrective action considering branch overloads and cut-set saturation (using generation rescheduling and load shed)
%     \item Our additions
%     \begin{itemize}
%         \item Add transient stability constraints (obtained from the previous section)
%         \item Modify optimization to include topology reconfiguration as well 
%     \end{itemize}
% \end{itemize}

% \vspace{-4pt}
\section{Results}\label{section4}

The proposed formulation is tested on the well-known IEEE 118-bus system.   
% \cite{pstca118bus}. 
% details which are given in \autoref{tab:sysdets}. 
The system has 54 generators, 186 transmission lines and 99 loads, with a total capacity of 9,966 MW.
Simulations and contingency analyses are performed using Siemens' $\mathrm{PSSE}^\circledR$ \cite{siemenspsse}. The optimization model is solved using $\mathrm{Gurobi}$ \cite{gurobi}, and the OPF is run using $\mathrm{Pandapower}$ \cite{pandapower}, an open-source package in Python.
% for power system static security analyses. 
All computations are done on a computer with an Intel Core (TM) i7-11800H CPU @2.3GHz with 16GB of RAM and an RTX 3070Ti GPU. 
% \begin{table}[ht]
% \centering
% \caption{Network details of the IEEE 118-bus system}
% \begin{tabular}{|c|c|}
% \hline
% \textbf{Elements}         & \textbf{118 Bus} \\ \hline
% Bus                       & 118              \\ \hline
% Synchronous generators    & 54               \\ \hline
% % Inverter based resources    & 0               \\ \hline
% Transmission lines        & 186              \\ \hline
% Tra`ormers              & 11              \\ \hline
% Total generation capacity & 9966 MW          \\ \hline
% \end{tabular}
% \label{tab:sysdets}
% \end{table}
% \begin{figure}[hb]
% \centering
% \includegraphics[width=0.48\textwidth]{images/boxplot_figure.pdf}
% \caption{Load variation data for five loads in the IEEE 118-bus system}
% \label{fig:load_var}
% \end{figure}
% \begin{figure*}
% \centering
% \includegraphics[width=\textwidth]{images/IEEE118BusSystemB.jpeg}
% \caption{Single line diagram of the IEEE 118 bus sytem}
% \label{fig:118_bus}
% \end{figure*}
% \begin{figure*}
% \centering
% \includegraphics[width=\textwidth]{images/CA BUS Schematic ALT-01.jpg}
% \caption{Single line diagram of the 240 bus sytem}
% \label{fig:240_bus}
% \end{figure*}

Contingency details for an identified corridor of the 118-bus system that is to be impacted by a progressing wildfire
% simulating an actual wildfire 
is given in \autoref{tab:cont_solns} (see first row).
When the fire does enter the security buffer \cite{8620983} of the two lines, a set of five faults is assumed to occur consecutively over a period of three seconds on both the lines, at the end of which the lines suffer a permanent outage. 
% A set of five arc-faults occur consecutively over a period of three seconds at the end of which each line suffers a permanent outage.
% A set of five arc-faults occur consecutively, with an interval
% % over a period 
% of 3 seconds in between,
% at the end of which each line suffers a permanent outage.
To vary the loads of the 118-bus system, we find typical load variations occurring in the loads of the publicly available 2000-bus Synthetic Texas system \cite{birchfield2017grid}, and then superimpose those variations on the 118-bus system loads using kernel density estimation.
% Load variation data is mapped from the publicly available 2000-bus Synthetic Texas system \cite{birchfield2017grid} using kernel density estimation. 
% The resulting variation for a few of the loads are shown in \autoref{fig:load_var}.
% Then, we sample from the resulting distribution 
% The kernel density estimates are then randomly sampled 
We generate $28,000$ samples from the resulting load values to create a dataset for training $\Upsilon$. 
This dataset
% The dataset derived from the KDE sampling 
captures the full spectrum of real-time load fluctuations and is representative of the operational variability of the loads for the time horizon considered.
% The corrective action analysis to generate the power shift required for all the loading conditions are then calculated using \eqref{eq:trans_stab_const}, and $\Upsilon$ is trained, with the train to test split in the ratio of $70:30$. 

% \begin{table}[ht]
% \centering
% \label{tab:cont_dets}
% \begin{tabular}{|c|c|c|l|}
% \hline
% \textbf{System} & \textbf{\begin{tabular}[c]{@{}c@{}}\# unstable\\ generators\end{tabular}} & \textbf{\begin{tabular}[c]{@{}c@{}}Generation\\  at risk\end{tabular}}  &\textbf{\begin{tabular}[c]{@{}c@{}}Revenue\\  loss(\$)\end{tabular}}\\ \hline
% 118 bus   & 2                                                                         &    534 MW                                                                     &12,651.44\\\hline
% \end{tabular}
% \end{table}

% \subsection{Contingency instability alleviation}
\subsection{Wildfire Contingency Impacts and TSCF Estimation Results}

\autoref{tab:cont_solns} shows the security and stability constraints that are obtained when the contingency mentioned in the first row occurs.
% generated. 
Note that the required rescheduling is considerable (second, fifth, and sixth rows).
% quite large. 
Furthermore, such a large change leads to additional static security violations (overloaded cut-sets), as seen in the third and fourth rows.
% for both the systems, 
% which are caught by the FT algorithm. 
% For these reasons, it is very essential that corrective action information is known beforehand, and is not entrusted to real time control strategies. 

% \begin{table}[ht]
% \centering
% \caption{Contingency details}
% \label{tab:cont_dets}
% \begin{tabular}{|c|c|c|c|}
% \hline
% \textbf{System} & \textbf{Lines tripped} & \textbf{\begin{tabular}[c]{@{}c@{}}\# unstable\\ generators\end{tabular}} & \textbf{\begin{tabular}[c]{@{}c@{}}Generation\\  at risk\end{tabular}} \\ \hline
% 118 bus   &     (30,26),(23,25)        & 2                                                                         &    534 MW                                                                    \\ \hline
% 240 bus   &   {\begin{tabular}[c]{@{}c@{}}(4102, 4202), (4202,4203), \\ (4202, 4005)\end{tabular}}                     & 7                                                                       &     6209.1 MW                                                                   \\ \hline
% \end{tabular}
% \end{table}

\begin{table}[ht]
\centering
\caption{Contingency details and alleviation actions}
\label{tab:cont_solns}
\begin{tabular}{|l|c|c|}
\hline
\textbf{Property}                                                                                & \textbf{Value} \\ \hline
Lines impacted &  (23,25),(26,30)\\ \hline

Generation at risk & 534 MW\\ \hline

{Saturated cut-sets }                                                                   &     (26-30,25-27)                     \\ \hline
{Cut-set transfer margin}                                                               &       -187.086 MW                 \\ \hline
% {\% of capacity}                                                                        &          42.51\%               \\ \hline
Critical machines (CMs) &   25,26 \\ \hline

% {\begin{tabular}[c]{@{}c@{}}Transient stability \\ critical machines (CM)\end{tabular}} &        25,26                 \\ \hline
{TSCF}                                                                                  &         -118 MW                \\ \hline
% {\% of capacity}                                                                        &          22.1\%               \\ \hline
\end{tabular}
\end{table}

% \begin{table}[ht]
% \centering
% \caption{Contingency alleviation actions}
% \begin{tabular}{|c|c|}
% \hline
% \textbf{Property}                                                                                & \textbf{Value}  \\ \hline

% N-1 branch overloads identified & (8-5)\\ \hline
% {Saturated cut-sets }                                                                   &     (26-30,25-27)                    \\ \hline
% {Cut-set transfer margin}                                                               &       -187.086 MW                 \\ \hline

% {\begin{tabular}[c]{@{}c@{}}Transient stability \\ critical machines (CM)\end{tabular}} &        25,26                  \\ \hline
% {TSCF}                                                                                  &         118 MW                \\ \hline
% \end{tabular}
% \end{table}

Using the contingency analysis and load variation information, the data-driven TSCP algorithm is trained for TSCF prediction in real-time. 
% The results for the TSCP obtained using various data-driven models are given in \autoref{tab:shallow_models}.
% In this study, the 
In \autoref{tab:shallow_models}, the TSCF estimate obtained
% $\Upsilon$ trained 
using LR is compared with the one obtained using
% other robust shallow networks, including 
% popular shallow machine learning models such as
% ridge regression, least absolute shrinkage and selection operator (LASSO) regression,
ridge regression, support vector regression (SVR), XGBoost, random forest (RF), elastic net, decision tree (DT), and least absolute shrinkage and selection operator (LASSO). 
% a 24-level decision tree (DT) with $9,400$ leaves, a random forest (RF) with $100$ trees,  and a support vector regressor (SVR) with radial basis function kernel, among other models. 
% Detailed results can be found in \autoref{tab:shallow_models}.
% The models are compared based on key performance metrics, including the root mean squared error (RMSE), and the $R^2$ score.
It is observed from the root mean squared error (RMSE) and $R^2$ scores that LR 
% gives very good results.
and ridge regression give the best estimates, while RF, elastic net, DT, and LASSO have comparatively poorer performance.
% , while many other models are able to estimate the correction factor
% % to an accuracy of 1 MW. 
% to within an error of 1 MW.
% % Since the final constraint is expected to add a factor of safety to this margin (which makes the system barely stable), this error is considered acceptable for estimation.
% % This is because SIME is built on a quasi-linear relationship as explained in Section \ref{Section 3}.

% A crucial test in TSCF prediction is to evaluate the average bias in the predictions,  termed the mean bias deviation (MBD). It represents if the trained model is generally under-predicting or over-predicting the TSCF, calculated as
Now,
% Additionally, 
since the loading conditions
% load forecasts 
are 
% naturally 
themselves subject to errors, the reliability of the estimation is investigated by calculating the $R^2$ robustness scores,
% through the robustness drop in $R^2$ scores, 
which is the change in $R^2$ scores when the error in input is increased. 
The results show that
% most of 
the models are generally
% capable of estimating the TSCF satisfactorily and are also 
resilient to errors in the loads
% load forecasts 
(which was capped at 5\%), with LR giving the most consistent outcomes.
Finally, \autoref{tab:shallow_models} shows the mean bias deviation (MBD) results, which represents the average bias in the predictions.
It is an indication of whether the trained model is under-estimating or over-estimating, and is calculated using:
% The consistency of estimation is realized by computing the mean bias deviation (MBD), which is a representation of the average bias in the model predictions, as seen in the equation below
\begin{equation}
    \mathrm{MBD} =  \frac{\sum_{k} (y - \hat{y})}{k}
\end{equation}

Ideally, MBD should be $0$.
% , to prevent bias. 
However, in our case, over-estimation of TSCF is preferred, i.e., a negative value of MBD is better. Again, it can be observed from \autoref{tab:shallow_models} that LR gives a very small negative value for MBD.
These results justify our assertion that a linear model is sufficient for TSCP.
% It is also important to also consider the robustness drop in $R^2$, since the load forecasts are naturally subject to errors. 
 
% This is because the data-driven techniques are learning a quasi-linear relationship (see Section \ref{Section 3}).

 % The shallow models combined with the interpretable nature of the constraints ensure transparency, i.e, should prevailing conditions shift, a utility operator possesses the discretion to intervene at any stage of the implementation, allowing the application of custom solutions derived from their experience.

\begin{table}[t]
% \vspace{-4pt}
\centering
\caption{TSCF estimation results using different models}
\label{tab:shallow_models}
\begin{tabular}{|l|c|c|c|c|c|c|c|c|c|}
\hline
\textbf{Model} & \textbf{RMSE} & \textbf{$R^2$}  & 
  \begin{tabular}{c}
    $R^2$  \\
    \textbf{Robustness}
\end{tabular} & \textbf{MBD}   \\
\hline
Linear Regression & 0.31 & 0.98 & 0.0024 & \(-0.57\mathrm{e}{-4}\)    \\
\hline
Ridge Regression & 0.31 & 0.98  & 0.0027 & \(-0.57\mathrm{e}{-4}\)    \\
\hline
SVR & 0.42 & 0.97  & 0.0028 & 0.02 \\
\hline
XGBoost & 0.94 & 0.85  & 0.0035 & -0.004   \\
\hline
Random Forest & 1.51 & 0.61  & 0.0031 & 0.005  \\
\hline
Elastic Net & 2.13 & 0.22 & -0.0003 & \(-0.93\mathrm{e}{-4}\)     \\
\hline

Decision Tree & 2.33 & 0.06  & -0.0051 & 0.002   \\
\hline
LASSO & 2.34 & 0.06  &  0.0001 &\(-0.28\mathrm{e}{-4}\)
   \\
 \hline

\end{tabular}
% \vspace{-4pt}
\end{table}

\subsection{CSCOPF Implementation Results}

\begin{figure}[ht]
 % \vspace{5em}
	\centering
	\includegraphics[width=0.485\textwidth]{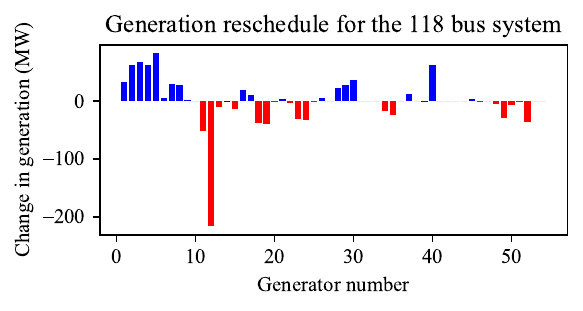}
    \vspace{-2em}
	\caption{CSCOPF redispatch schedule}
	\label{fig:118_genchange}
\end{figure}

% % Upon manifestation of the wildfire in real-time,
% When 
% % the wildfire manifests, and 
% the projected path of an ongoing wildfire is expected to impact
% % may affect 
% the region in operation,
% % enters the security buffer of the power lines,
% % risk is high,
% % (wildfire, in the context of this paper),
% the proposed CSCOPF is used to quickly redispatch the system.
The proposed CSCOPF is used to quickly redispatch the 118-bus system when the projected path of the wildfire is expected to impact 
% a region inside the system.
the region identified by the first row of \autoref{tab:cont_solns}.
The generator redispatch is shown in Fig. \ref{fig:118_genchange}. 
In the figure, a positive (blue) bar indicates increase in the generator's output, while a negative (red) bar indicates the opposite. 
The majority of generation shed occurs from
% is from 
the CMs identified in \autoref{tab:cont_solns}.
% to make the system transient stable \eqref{eq:trans_stab_const}. 
The FT algorithm and other security criteria ensure that this rescheduling does not increase vulnerability in other areas of the system (shown in subsequent results).

\vspace{-1em}

\begin{figure}
\centering
\begin{subfigure}{.425\textwidth}
  \centering
  \includegraphics[width=\textwidth]{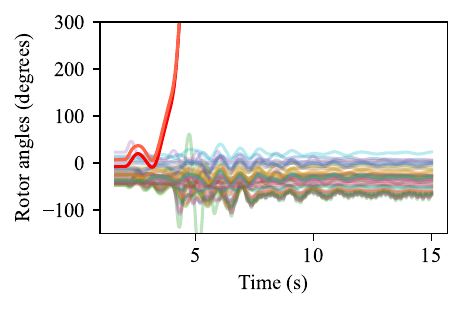}
  \vspace{-2em}
  \caption{TDS without CSCOPF}
  \label{fig:sub1}
\end{subfigure}\\
\begin{subfigure}{.425\textwidth}
  \centering
 \includegraphics[width=\textwidth]{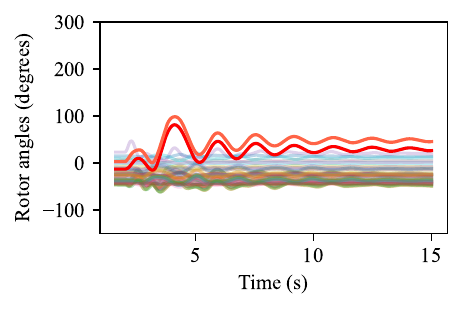}
  \vspace{-2em}
  \caption{TDS with CSCOPF}
  \label{fig:sub2}
\end{subfigure}
% \vspace{-1em}
\caption{Rotor angle stability for the IEEE 118-bus system}
\label{fig:118_rotorangles}
\end{figure}

A contingency analysis was done before and after implementation of the CSCOPF (Algorithm \ref{alg:cap2}). Rotor angle trajectories for both cases are shown in Fig. \ref{fig:118_rotorangles}. In the figure, the two CMs are highlighted in red and orange colors. 
% Without any control, the generators lose synchronism and diverge from the rest of the machines, which is rectified when CSCOPF is implemented. 
Without any control, the CMs diverged and quickly lost synchronism. With CSCOPF, they remained synchronized with the rest of the system, effectively alleviating the transient instabilities.
% and it show that the generators are able to ride-through the contingencies without losing synchronism. 

% Static security results for CSCOPF is 
% The final results obtained using CSCOPF are
% given in \autoref{tab:preventive_all}. In the table, the proposed algorithm is compared 
\autoref{tab:preventive_all} compares CSCOPF with 
% two other control approaches: 
a conventional real-time security constrained economic dispatch (RT-SCED), and a transient stability constrained OPF (TSCOPF).
% , both of which have been used to tackle extreme event scenarios.
% which is widely researched as a potential solution to extreme event scenarios. 
The RT-SCED does not have cut-set or stability constraints while the TSCOPF solution does not include cut-set constraints and performs SIME in real-time.
% The formulation of TSCOPF does not include the cut-set constraints. 
During implementation, the RT-SCED did not reschedule sufficient generation for ensuring transient stability as it did not have any dynamic information (power from CMs was only lowered by 20 MW as compared to the minimum needed 118 MW shown in \autoref{tab:cont_solns}). 
Moreover, in the identified cut-set, the power flow actually \textit{increased} instead of decreasing, making that region of the network even more vulnerable to cascading line outages. 
In comparison, TSCOPF successfully alleviated the transient instabilities, but could not fully address the cut-set overloads (cut-set transfer only reduced by 57 MW as compared to the minimum needed 186 MW shown in \autoref{tab:cont_solns}). This meant that the system was still at risk of cut-set saturation and line trips.
% In comparison, TSCOPF is able to expectedly alleviate the system of the transient instability, and it is able to reduce the cut-set saturation as well, however it does not completely alleviate it (cut-set transfer only reduced by 57 MW as compared to the minimum 186 MW), which means that the system is still at risk of cut-set saturation and line trips.

\begin{table}[hb]
\centering
% \caption{CSCOPF control results and their comparison with other approaches}
\caption{Comparative analysis of CSCOPF results}
\begin{tabular}{|l|c|c|c|c|c|}
\hline
\textbf{Result}                             & \textbf{RT-SCED} & \textbf{TSCOPF}& \textbf{CSCOPF} \\ \hline
CM generation shed (MW)          & 20.4 & 118 & 269.79 \\ \hline
% NM generation addition (MW) & & && 269.79 \\ \hline
Cut-set desaturation (MW) & -131.80 & 57.519 & 187.087\\ \hline
Total load shed (MW)                   &0 &0& 0 \\ \hline
Transient stable          & No & Yes & Yes \\ \hline
Cut-set secure & No & No & Yes\\ \hline
Time to solve (s)        & 0.066 & 0.256 & 0.066 \\ \hline
Cost (\$/hr)                  & 126,459.97 & 126,222.28 & 128,187.95           \\ \hline
\end{tabular}
% \newline
% \vspace*{5pt}
% \newline
% \begin{tabular}{|c|c|}
% \hline
%      \textbf{Model} & \textbf{Description} \\ \hline
%      A1 & Security Constrained Economic Dispatch  (SCED)\\ \hline
%      A2 & Transient Stability Constrained OPF (TSCOPF) \\ \hline
%      A3 & Proposed CSCOPF \\

%  \hline
% \end{tabular}

\label{tab:preventive_all}
\end{table}

The proposed CSCOPF alleviated both static and dynamic vulnerabilities resulting in a resilient operation without incurring any load shed (see last column of \autoref{tab:preventive_all}).
The solution time of CSCOPF is similar to RT-SCED and one-fourth of TSCOPF, implying that the additional constraints do not increase the computational burden of the optimization.
% , with the data-driven transient stability analyzer actually improving the speed of online operation.
% The optimization time of CSCOPF is also similar to the other approaches, implying that the additional constraints do not significantly increase the computational burden of the optimization.
From an economic perspective, CSCOPF incurred an additional cost of \$1728/hr (over the RT-SCED solution). For context, in absence of any control, the simulated wildfire-induced contingency
% a system with no control, 
% an insecure solution 
would have put about 534 MW of generation at risk of tripping (see \autoref{tab:cont_solns}), which would have led to a loss in revenue of at least \$12,650/hr calculated on the generation side.
% Meanwhile the proposed CSCOPF is able to alleviate both vulnerabilities, yielding in an actual secure solution without any load shed. The optimization time of CSCOPF is also similar to the other models, which means that the additional stability does not incur any computational burden. Economically, CSCOPF has an additional operational cost of about \$ 1727.48. For context, a system with no control/an insecure solution would have put about 534 MW of generation at risk of tripping which would have led to a loss in revenue of at least \$12,651.44, calculated on the generation side.
Lastly, although load-shed was not required for the contingency-under-study, it is important to incorporate it in the problem  formulation as a different contingency may require both generation redispatch as well as load-shed.

% base cost = 126,222.28
% \textcolor{red}{Show line voltages and show that no more cascading line outages, maybe even try to show line outages before}.

% \subsection{Economic Analysis}
% The cost of operation of the proposed CSCOPF is shown in \textcolor{red}{Table}. The additional cost of secure operation is compared with the potential loss of generation when the contingency hits an unsecured system. The potential loss of load shed outweighs the additional cost of secure operation.

% \vspace{-2pt}
\section{Conclusion}\label{section5}

In this paper, a comprehensive corrective action scheme addressing both static and dynamic insecurities
% security criteria 
was introduced to ensure
% that resulted in a 
resilient power system operation
% security and stability-constrained optimal power flow solutions 
during active wildfires. 
% We further proposed a 
% Considering the unique effects of extreme events on power systems, 
The scheme is based on an advanced contingency analysis tool that accurately analyses the impacts of such extreme event scenarios. 
% An advanced contingency analysis method is developed to accurately analyse the impacts of extreme event contingency scenarios. 
The tool includes: (i) a static security component that is able to exhaustively desaturate cut-sets and prevent cascading line outages, thereby going beyond traditional security approaches that only protect against branch overloads, and
% clear cut-set saturation 
% is implemented
% It goes 
% covering 
% beyond traditional security approaches that only protect against branch overloads.
% N-1 security risks.
% It also includes 
(ii) a data-driven
% linear 
TSCP model
that 
% able to 
accurately and reliably predicts the appropriate correction factor for mitigating transient instabilities under varying loading conditions and prevents cascade tripping of generators. 
% The LR-based TSCP algorithm with convex constraints
Finally, the 
% Moreover, an 
LR-based implementation guarantees both transparency (in comparison to black-box models) as well as solution optimality.
% The shallow TSCP algorithm combined with the convex constraints ensure transparency, i.e, should prevailing conditions shift, a utility operator possesses the discretion to intervene at any stage of the implementation. 

The numerical results indicate that the proposed model is able to detect and alleviate cascading outage risks due to overloaded lines, generators, as well as cut-sets,
% of both transmission lines and generators, 
while bearing minimal additional operational cost. 
% Future work will analyse the impact of renewable resources on CSCOPF from both economic as well as stability perspectives.
Future work will analyse the ability of the proposed approach in mitigating the impacts of wildfires on renewable-rich systems from both stability as well as economic perspectives.

\bibliography{bibtex.bib}

% Generated by IEEEtran.bst, version: 1.14 (2015/08/26)
\def\authornoop#1{}
\begin{thebibliography}{10}
\providecommand{\url}[1]{#1}
\csname url@samestyle\endcsname
\providecommand{\newblock}{\relax}
\providecommand{\bibinfo}[2]{#2}
\providecommand{\BIBentrySTDinterwordspacing}{\spaceskip=0pt\relax}
\providecommand{\BIBentryALTinterwordstretchfactor}{4}
\providecommand{\BIBentryALTinterwordspacing}{\spaceskip=\fontdimen2\font plus
\BIBentryALTinterwordstretchfactor\fontdimen3\font minus \fontdimen4\font\relax}
\providecommand{\BIBforeignlanguage}[2]{{%
\expandafter\ifx\csname l@#1\endcsname\relax
\typeout{** WARNING: IEEEtran.bst: No hyphenation pattern has been}%
\typeout{** loaded for the language `#1'. Using the pattern for}%
\typeout{** the default language instead.}%
\else
\language=\csname l@#1\endcsname
\fi
#2}}
\providecommand{\BIBdecl}{\relax}
\BIBdecl

\bibitem{balaraman2020wildfires}
K.~Balaraman, ``{Wildfires pushed PG\&E into bankruptcy. Should other utilities be worried?}'' \emph{Utility Dive}, Nov. 19, 2020, \url{https://www.utilitydive.com/news/wildfires-pushed-pge-into-bankruptcy-should-other-utilities-be-worried/588435/}.

\bibitem{9737413}
P.~Moutis and U.~Sriram, ``{PMU}-driven non-preemptive disconnection of overhead lines at the approach or break-out of forest fires,'' \emph{IEEE Transactions on Power Systems}, vol.~38, no.~1, pp. 168--176, 2023.

\bibitem{wu2016space}
Y.~Wu, Y.~Xue, J.~Lu, Y.~Xie, T.~Xu, W.~Li, and C.~Wu, ``Space-time impact of forest fire on power grid fault probability,'' \emph{Automation of Electric Power Systems}, vol.~40, no.~3, pp. 14--20, 2016.

\bibitem{9677975}
D.~A.~Z. Vazquez, F.~Qiu, N.~Fan, and K.~Sharp, ``Wildfire mitigation plans in power systems: A literature review,'' \emph{IEEE Transactions on Power Systems}, vol.~37, no.~5, pp. 3540--3551, 2022.

\bibitem{CHOOBINEH201520}
M.~Choobineh, B.~Ansari, and S.~Mohagheghi, ``Vulnerability assessment of the power grid against progressing wildfires,'' \emph{Fire Safety Journal}, vol.~73, pp. 20--28, 2015.

\bibitem{nerc1200}
NERC, ``1,200 {MW} fault induced solar photovoltaic resource interruption disturbance report: Southern {C}alifornia 8/16/2016 event,'' \url{https://www.nerc.com/pa/rrm/ea/Pages/1200-MW-Fault-Induced-Solar-Photovoltaic-Resource-Interruption-Disturbance-Report.aspx}.

\bibitem{8620983}
S.~Dian, P.~Cheng, Q.~Ye, J.~Wu, R.~Luo, C.~Wang, D.~Hui, N.~Zhou, D.~Zou, Q.~Yu, and X.~Gong, ``Integrating wildfires propagation prediction into early warning of electrical transmission line outages,'' \emph{IEEE Access}, vol.~7, pp. 27\,586--27\,603, 2019.

\bibitem{daochun2015review}
H.~Daochun, L.~Peng, R.~Jiangjun, Z.~Yafei, and W.~Tian, ``Review on discharge mechanism and breakdown characteristics of transmission line gap under forest fire condition,'' \emph{High Voltage Engineering}, vol.~41, no.~2, pp. 622--632, 2015.

\bibitem{9449670}
C.~Haseltine and L.~Roald, ``The effect of blocking automatic reclosing on wildfire risk and outage times,'' in \emph{2020 52nd North American Power Symposium}, 2021, pp. 1--6.

\bibitem{Columbia}
J.~J. Macwilliams, J.~Kobus, and S.~L. Monaca, ``{PG\&E}: Market and policy perspectives on the first climate change bankruptcy,'' 2019.

\bibitem{shrestha2021frequency}
\BIBentryALTinterwordspacing
A.~Shrestha and F.~Gonzalez-Longatt, ``Frequency stability issues and research opportunities in converter dominated power system,'' \emph{Energies}, vol.~14, no.~14, 2021. [Online]. Available: \url{https://www.mdpi.com/1996-1073/14/14/4184}
\BIBentrySTDinterwordspacing

\bibitem{biswas2020graph}
R.~S. Biswas, A.~Pal, T.~Werho, and V.~Vittal, ``A graph theoretic approach to power system vulnerability identification,'' \emph{IEEE Transactions on Power Systems}, vol.~36, no.~2, pp. 923--935, 2020.

\bibitem{biswas2021mitigation}
{\authornoop{I}}{R. S. Biswas, A. Pal, T. Werho, and V. Vittal}, ``Mitigation of saturated cut-sets during multiple outages to enhance power system security,'' \emph{IEEE Transactions on Power Systems}, vol.~36, no.~6, pp. 5734--5745, 2021.

\bibitem{ma2020real}
J.~Ma, J.~C. Cheng, F.~Jiang, V.~J. Gan, M.~Wang, and C.~Zhai, ``Real-time detection of wildfire risk caused by powerline vegetation faults using advanced machine learning techniques,'' \emph{Advanced Engineering Informatics}, vol.~44, p. 101070, 2020.

\bibitem{flammap}
``{FlamMap},'' {US Forest Service, Rocky Mountain Research Station, Fire, Fuel, and Smoke Science Program}, \url{https://www.firelab.org/project/flammap}, last accessed October 2023.

\bibitem{9693217}
M.~Abdelmalak and M.~Benidris, ``Enhancing power system operational resilience against wildfires,'' \emph{IEEE Transactions on Industry Applications}, vol.~58, no.~2, pp. 1611--1621, 2022.

\bibitem{8244267}
J.~Lu, J.~Guo, Z.~Jian, and X.~Xu, ``Optimal allocation of fire extinguishing equipment for a power grid under widespread fire disasters,'' \emph{IEEE Access}, vol.~6, pp. 6382--6389, 2018.

\bibitem{9409078}
A.~Arab, A.~Khodaei, R.~Eskandarpour, M.~P. Thompson, and Y.~Wei, ``Three lines of defense for wildfire risk management in electric power grids: A review,'' \emph{IEEE Access}, vol.~9, pp. 61\,577--61\,593, 2021.

\bibitem{10032578}
R.~Bayani and S.~D. Manshadi, ``Resilient expansion planning of electricity grid under prolonged wildfire risk,'' \emph{IEEE Transactions on Smart Grid}, vol.~14, no.~5, pp. 3719--3731, 2023.

\bibitem{sahoo2023data}
S.~Sahoo, A.~I. Sifat, and A.~Pal, ``Data-driven flow and injection estimation in {PMU}-unobservable transmission systems,'' in \emph{2023 IEEE Power \& Energy Society General Meeting}.\hskip 1em plus 0.5em minus 0.4em\relax IEEE, 2023, pp. 1--5.

\bibitem{9290096}
C.~Guo, C.~Ye, Y.~Ding, and P.~Wang, ``A multi-state model for transmission system resilience enhancement against short-circuit faults caused by extreme weather events,'' \emph{IEEE Transactions on Power Delivery}, vol.~36, no.~4, pp. 2374--2385, 2021.

\bibitem{zhang1998sime}
Y.~Zhang, L.~Wehenkel, and M.~Pavella, ``{SIME}: A comprehensive approach to fast transient stability assessment,'' \emph{IEEJ Transactions on Power and Energy}, vol. 118, no.~2, pp. 127--132, 1998.

\bibitem{7395386}
Y.~Xu, J.~Ma, Z.~Y. Dong, and D.~J. Hill, ``Robust transient stability-constrained optimal power flow with uncertain dynamic loads,'' \emph{IEEE Transactions on Smart Grid}, vol.~8, no.~4, pp. 1911--1921, 2017.

\bibitem{siemenspsse}
\BIBentryALTinterwordspacing
{Siemens Industry Software Inc.}, \emph{PSS\textsuperscript{\textregistered}E}, Siemens, Washington, DC, 2023, version 35.0.0. [Online]. Available: \url{https://www.siemens.com/global/en/products/energy/grid-software/planning/pss-software/pss-e.html}
\BIBentrySTDinterwordspacing

\bibitem{gurobi}
\BIBentryALTinterwordspacing
L.~Gurobi~Optimization, \emph{Gurobi Optimizer Reference Manual}, Gurobi Optimization, LLC, 2023, version 10.0. [Online]. Available: \url{https://www.gurobi.com}
\BIBentrySTDinterwordspacing

\bibitem{pandapower}
\BIBentryALTinterwordspacing
L.~Thurner, A.~Scheidler, F.~Schäfer, J.-H. Menke, J.~Dollichon, F.~Meier, S.~Meinecke, and M.~Braun, ``Pandapower,'' 2022, version 2.9.0. [Online]. Available: \url{http://www.pandapower.org}
\BIBentrySTDinterwordspacing

\bibitem{birchfield2017grid}
A.~B. Birchfield, T.~Xu, K.~M. Gegner, K.~S. Shetye, and T.~J. Overbye, ``Grid structural characteristics as validation criteria for synthetic networks,'' \emph{IEEE Transactions on Power Systems}, vol.~32, no.~4, pp. 3258--3265, 2017.

\end{thebibliography}

\end{document}